\documentclass[conference]{IEEEtran}
\IEEEoverridecommandlockouts                              

\usepackage[letterpaper, left=0.75in, right=0.75in, bottom=0.8in, top=0.75in]{geometry}

\usepackage{amsmath} 
\usepackage{amssymb}  
\usepackage{todonotes}
\usepackage{enumerate}
\usepackage{float}
\usepackage{subcaption}
\usepackage{afterpage}

\newcommand\R{\mathbb{R}}
\newtheorem{theorem}{Theorem}

\newtheorem{definition}{Definition}

\title{\LARGE \bf
Non-Submodular Maximization via the Greedy Algorithm and the Effects of Limited Information in Multi-Agent Execution
}

\author{\IEEEauthorblockN{Benjamin Biggs}
\IEEEauthorblockA{\textit{Virginia Tech}\\
babiggs@vt.edu}
\and
\IEEEauthorblockN{James McMahon}
\IEEEauthorblockA{\textit{US Naval Research Laboratory}\\ Acoustics Division, Code 7130 \\
    james.mcmahon@nrl.navy.mil}
\and
\IEEEauthorblockN{Philip Baldoni}
\IEEEauthorblockA{\textit{US Naval Research Laboratory}\\ Acoustics Division, Code 7130 \\
    philip.baldoni@nrl.navy.mil}
\and 
\IEEEauthorblockN{Daniel J. Stilwell}
\IEEEauthorblockA{\textit{Virginia Tech}\\
stilwell@vt.edu}
\thanks{*This work was supported by the Office of Naval Research via grants N00014-18-1-2627, and N00014-19-1-2194. The work of J. McMahon and P. Baldoni is supported by the Office of Naval Research through the NRL Base Program.}
\thanks{James McMahon and Phillip Baldoni are with the US Naval Research Laboratory, Code 7130, Washington D.C., USA}%
\thanks{Benjamin Biggs and Daniel Stilwell are with the Bradley Department of Electrical and Computer Engineering, Virginia Tech, Blacksburg, VA, USA}%

}

\begin{document}
\maketitle
\thispagestyle{empty}
\pagestyle{empty}

\begin{abstract}
    We provide theoretical bounds on the worst case performance of the greedy algorithm in seeking to maximize a normalized, monotone, but not necessarily submodular objective function under a simple partition matroid constraint. We also provide worst case bounds on the performance of the greedy algorithm in the case that limited information is available at each planning step. We specifically consider limited information as a result of unreliable communications during distributed execution of the greedy algorithm. We utilize notions of curvature for normalized, monotone set functions to develop the bounds provided in this work. To demonstrate the value of the bounds provided in this work, we analyze a variant of the benefit of search objective function and show, using real-world data collected by an autonomous underwater vehicle, that theoretical approximation guarantees are achieved despite non-submodularity of the objective function.
\end{abstract}

\section{INTRODUCTION}
In general, the problem of planning search paths that seek to maximize a general objective function is NP-hard. One simple method of addressing the general infeasibility of planning optimal paths for a team of search agents is to utilize the \textit{greedy algorithm} \cite{conforti1984submodular} wherein an ordering is assigned to the set of search agents and each agent plans a path for itself while accounting for the paths of preceding agents. We seek to provide theoretical approximation guarantees for the greedy algorithm in seeking to maximize a non-submodular objective function where planning at each step of the greedy algorithm is potentially suboptimal. Furthermore, we consider the effect of unreliable communications during distributed execution of the greedy algorithm which limits the information available to search agents and seek to provide approximation guarantees for the greedy algorithm with limited information as well.

The greedy algorithm has received significant attention because of its practical simplicity. In addition to its ease of implementation, the greedy algorithm has been shown to yield results with high-quality approximation guarantees under a wide variety of constraints on the possible solutions that the greedy algorithm may produce. 

It is shown in \cite{edmonds1971matroids} that the greedy algorithm yields an optimal solution given a modular objective function. The well known $1/2$ lower bound is presented in \cite{fisher1978analysis} for the case of a normalized, monotone, submodular objective function under a general matroid constraint. The authors of \cite{conforti1984submodular} introduce a notion of curvature for normalized, monotone, submodular objective functions and use this curvature $\alpha_c \in [0,1]$ to provide an approximation guarantee of $1/(1 + \alpha_c)$ that is equivalent to the bounds presented in \cite{edmonds1971matroids} and \cite{fisher1978analysis} at the limits of the curvature term (when $\alpha_c = 0$ or $\alpha_c = 1$) thus unifying the results of \cite{edmonds1971matroids} and \cite{fisher1978analysis}. It is shown in \cite{nemhauser1978analysis} that under a uniform matroid constraint, the greedy algorithm yields an improved bound of $(1 - e^{-1}) \approx 0.63$ given a submodular objective function. In \cite{conforti1984submodular}, curvature is again used to further improve this bound to $(1 - e^{-\alpha_c})/\alpha_c$.

The bounds of \cite{fisher1978analysis} and \cite{nemhauser1978analysis} have motivated the frequent use of submodular objective functions such as mutual information which is known to be submodular assuming measurements are conditionally independent \cite{krause2012near}. Further consideration of approximation guarantees for the greedy algorithm in seeking to maximize a non-submodular objective function has begun only recently. We refer the reader to \cite{lu2022maximizing} for a more thorough discussion of distinct types of non-submodular functions and recent advancements. Important applications with corresponding non-submodular objective functions include experimental design, dictionary selection, and subset selection \cite{lu2022maximizing}. The work of \cite{gatmiry2021network} addresses visibility optimization in social media. In this work, we address robotic search.

\textbf{Contributions:} In the first contribution of this work, which appears in Theorem \ref{thm.greedy_bound}, we provide suboptimality guarantees for the greedy algorithm under a simple partition matroid constraint where the objective function is normalized and monotone, but not necessarily submodular. We discuss how our novel bound generalizes and extends bounds provided in earlier works with the additional consideration of generalized curvature. In the second contribution of our work, which appears in Theorem \ref{thm.lim_inf_greedy_bound}, we provide suboptimality guarantees for the greedy algorithm with limited information extending the work of \cite{gharesifard2017distributed, grimsman2018impact}. Our extensions include the consideration of normalized, monotone, but not necessarily submodular objective functions. Additionally, in each case we consider that the greedy selection step is executed with bounded suboptimality as in \cite{singh2009efficient}. Finally, we illustrate the efficacy of our contributions with respect to a robotic search application.

Most closely related to the first contribution of this work are the results presented in \cite{nong2019maximize} and \cite{gatmiry2021network} which show that under a general matroid constraint the greedy algorithm yields a constant approximation factor $(1 - \beta)/(1 + (1 - \beta))$ for normalized, monotone, but not necessarily submodular objective functions where $\beta$ is equivalent to the inverse generalized curvature given in Definition \ref{def.bogunovic_inverse_curvature} of our work. Our work extends the bounds of \cite{nong2019maximize, gatmiry2021network} by further considering the generalized curvature given in Definition \ref{def.bogunovic_curvature}. Our contribution given as Theorem \ref{thm.greedy_bound} has a form very similar to that of the bound presented in \cite{bai2018greed}. In contrast to \cite{bai2018greed} where the authors consider the sum of submodular and supermodular functions and utilize notions of curvature with respect to these functions, we consider a general non-submodular objective function. This is significant because a decomposition of a normalized monotone set function into a sum of submodular and supermodular functions is not guaranteed to exist as shown in Lemma 3.2 of \cite{bai2018greed}. Additionally, we consider that the greedy selection step is potentially suboptimal as in \cite{singh2009efficient} further distinguishing our contribution.

Most closely related to the second contribution of this work are the results in \cite{gharesifard2017distributed} and \cite{grimsman2018impact} where a generalized greedy algorithm is presented that does not require complete information regarding preceding decisions. Notably, the generalized greedy algorithm in \cite{grimsman2018impact} enjoys a worst-case approximation guarantee of $1/(1 + k^*(G))$ under a partition matroid constraint assuming a normalized, monotone, submodular objective function. The term $k^*(G)$ represents the fractional clique cover number of the underlying communication graph $G$ and is discussed in detail in \cite{grimsman2018impact}. We extend the work of \cite{grimsman2018impact} by considering the curvature of the objective function which is assumed to be normalized and monotone, but not necessarily submodular. 

\textbf{Paper Overview:} This paper is organized as follows. Mathematical preliminaries are presented and discussed in Section \ref{sec.preliminaries}. Properties of the objective function including notions of curvature are presented in Section \ref{sec.properties_of_objective}. The primary contributions of this work are presented in Section \ref{sec.guarantees}. The benefit of search objective function is presented in Section \ref{sec.benefit_of_search} and properties of the objective function are analyzed in Section \ref{sec.benefit_of_search_properties}. Conclusions are presented in Section \ref{sec.conclusions}. Proofs are presented in the appendices.

\section{Mathematical Preliminaries}\label{sec.preliminaries}
Let $\mathcal{V}$ be a finite ground set and let $f: 2^{\mathcal{V}} \mapsto \R$ be a set function where $2^{\mathcal{V}}$ denotes the power set of $\mathcal{V}$. Generally, calligraphic font is used to denote sets. Exceptions should be clear from context. We consider a team of $N$ agents given as $\mathcal{A} = \{1,\ldots, N \}$. In executing the greedy algorithm, each agent $i$ will select an element $x_i$ from the ground set $\mathcal{V}$ in an effort to maximize the objective function $f$. 

When analyzing suboptimality of the greedy algorithm, one often encounters matroids which are used to represent constraints on the solutions produced by the greedy algorithm. A matroid is defined as follows.
\begin{definition}[Matroid {\cite[Definition 39.1]{schrijver2003combinatorial}} ]
    Consider a finite ground set $\mathcal{V}$, and a non-empty collection of subsets of $\mathcal{V}$, denoted by $\mathcal{I}$. Then, the pair $(\mathcal{V}, \mathcal{I})$ is called a matroid if and only if the following conditions hold:
    \begin{enumerate}[i )]
        \item for any set $\mathcal{X} \subseteq \mathcal{V}$ such that $\mathcal{X} \in \mathcal{I}$, and for any set $\mathcal{Z} \subseteq \mathcal{X}$, it holds $\mathcal{Z} \in \mathcal{I}$;
        \item for any sets $\mathcal{X}, \mathcal{Z} \subseteq \mathcal{V}$ such that $\mathcal{X}, \mathcal{Z} \in \mathcal{I}$ and $|\mathcal{X}| < |\mathcal{Z}|$, it holds that there exists an element $z \in \mathcal{Z} \setminus \mathcal{X}$ such that $\mathcal{X}
        \cup \{ z\} \in \mathcal{I}$.
    \end{enumerate}
\end{definition}
Matroids are used to represent abstract dependence. The set $\mathcal{I}$ contains all independent sets of $\mathcal{V}$. A set $x \in \mathcal{I}$ is called maximal if there exists no $v \in \mathcal{V}$ such that $x \cup \{ v \} \in \mathcal{I}$.  For example, any subset of the columns of a matrix are either linearly independent or dependent. A matroid may be constructed by allowing the columns of the matrix to form the ground set $\mathcal{V}$ and every combination of linearly independent columns to form $\mathcal{I}$. An excellent discussion of matroids and their history is provided in \cite{schrijver2003combinatorial}. Most importantly for our purposes, matroids provide a construct to represent constraints in planning. Two notable special types of matroids are uniform matroids \cite{conforti1984submodular, nemhauser1978analysis, bian2017guarantees} and partition matroids \cite{grimsman2018impact, singh2009efficient, corah2018distributed, corah2019distributed, friedrich2019greedy}.

To analyze suboptimality bounds for the greedy algorithm, we utilize the \textit{marginal reward}.
\begin{definition}[Marginal Reward]\label{def.marginal_reward}
    Given any sets $\mathcal{S}, \mathcal{Q} \subseteq \mathcal{V}$ the marginal contribution of $\mathcal{S}$ given $\mathcal{Q}$ is 
\begin{equation} \label{eq.marginal_reward}
    \Delta(\mathcal{S}| \mathcal{Q}) \triangleq f(\mathcal{S} \cup \mathcal{Q}) - f(\mathcal{Q}).
\end{equation}
\end{definition}
We adopt notation for the marginal reward from \cite{gharesifard2017distributed} and \cite{grimsman2018impact}. It may be useful to consider the marginal reward as the discrete derivative of $f$ as discussed in \cite{corah2018distributed}.

Note that given an arbitrarily ordered set $\mathcal{S} \subseteq \mathcal{V}$, the reward attained by $\mathcal{S}$ may be represented as the sum of the marginal rewards for the elements of $\mathcal{S}$. For example, given $\mathcal{S} = \{ s_1, \ldots, s_M\}$, we have
\begin{equation}\label{eq.sum_of_marginals}
    f(\mathcal{S}) = \sum_{i = 1}^M \Delta(\{ s_i \} | \cup_{j = 1}^{i-1} \{s_j \})
\end{equation}
where $\cup_{j = 1}^{i-1} \{s_j \}$ is the empty set when $i = 1$. For convenience, we use the subscript notation $s_{1:i-1} = \cup_{j = 1}^{i-1} \{s_j \}$ to denote a subset of an ordered set. We also often use a comma in place of the union operator, e.g. $\Delta(\mathcal{S} | \mathcal{Q}, \mathcal{Z}) = \Delta(\mathcal{S} | \mathcal{Q} \cup \mathcal{Z})$. Lastly, when representing a single element in a set, we often drop the curly brace notation, e.g. $\Delta(\{ s_i \} | \mathcal{Q}) = \Delta(s_i | \mathcal{Q})$.

By \eqref{eq.sum_of_marginals}, given sets $\mathcal{S}, \mathcal{Q} \subseteq \mathcal{V}$ with $|\mathcal{S}| = M_\mathcal{S}$ and $|\mathcal{Q}| = M_\mathcal{Q}$ and arbitrary orderings $\mathcal{S} = \{ s_1, \ldots, s_{ M_\mathcal{S}}\}$ and $\mathcal{Q} = \{ q_1, \ldots, q_{ M_\mathcal{Q}}\}$ we have
\begin{align}\label{eq.reward_of_union}
    f(\mathcal{S}, \mathcal{Q}) &= f(\mathcal{S}) + \sum_{i = 1}^{M_\mathcal{Q}} \Delta(q_i | q_{1:i-1}, \mathcal{S}) \\ 
    f(\mathcal{S}, \mathcal{Q}) &= f(\mathcal{Q}) + \sum_{i = 1}^{M_\mathcal{S}} \Delta(s_i | s_{1:i-1}, \mathcal{Q}).
\end{align}

For the case of informative path planning, let $\mathcal{X}_i$ represents the set of paths available to agent $i$ and let $\mathcal{V} = \cup_{i \in \mathcal{A}}\mathcal{X}_i$. We define the constraint $\mathcal{I} = \{ \mathcal{S} \subseteq \mathcal{V} : |\mathcal{S} \cap \mathcal{X}_i| \leq 1 \ \forall i \in \mathcal{A}\}$, which is a simple partition matroid. Any element of $\mathcal{I}$ gives a joint path for a set of agents $a \subseteq \mathcal{A}$ and maximal set in $\mathcal{I}$ gives a complete joint path. That is, a maximal set $x$ includes an assignment for each agent $i$ from its set of paths $\mathcal{X}_i$. Let $x,x^* \in \mathcal{I}$ be maximal independent sets in $\mathcal{I}$. Then replacing $\mathcal{S}$ and $\mathcal{Q}$ in \eqref{eq.reward_of_union} with $x$ and $x^*$ gives us
\begin{align}
    f(x, x^*) &= f(x^*) + \sum_{i \in \mathcal{A}} \Delta(x_i | x_{1:i - 1}, x^*) \\
    f(x, x^*) &= f(x) + \sum_{i \in \mathcal{A}} \Delta(x_i^* | x_{1:i - 1}^*, x)
\end{align}
which, in turn, gives
\begin{equation}\label{eq.start_equation}
    f(x^*) = f(x) + \sum_{i \in \mathcal{A}} \Delta(x_i^* | x_{1:i - 1}^*, x) - \sum_{i \in \mathcal{A}} \Delta(x_i | x_{1:i - 1}, x^*).
\end{equation}
While equality in \eqref{eq.start_equation} holds for any maximal $x, x^* \in \mathcal{I}$, $x^*$ is assumed to satisfy $x^* \in \underset{\mathcal{S} \in \mathcal{I}, |\mathcal{S}| = N}{\arg \max} f(\mathcal{S})$ while $x$ is assumed to be a solution produced by the greedy algorithm.

In addition, we consider the case that limited information is available to an agent at its planning turn as in \cite{gharesifard2017distributed, grimsman2018impact}. More precisely, the standard greedy algorithm relies on the plans of agents $1$ through $i-1$ to be available to agent $i$ when it plans. We consider that only a subset of those plans $\mathcal{N}_i \subseteq \{j \in \mathcal{A} : j < i \}$ is available to agent $i$ when planning. Thus, for informative path planning, we denote the incomplete joint plan available to agent $i$ with $x_{\mathcal{N}_i}$. As a note: consider that each element of a set $\mathcal{S} \subseteq \mathcal{V}$ is represented by a node in a graph $G$ and let the in-neighbors of each node $s_i$ be the set of nodes $ \{s_j : 1 \leq j < i \}$ such that the $G$ is a directed acyclic graph. Equality in \eqref{eq.sum_of_marginals} is only guaranteed when $G$ is complete, i.e. $\mathcal{N}_i = \{j \in \mathcal{A} : j < i \}$. Specifically for a subset $\mathcal{S}$ of nodes, \eqref{eq.sum_of_marginals} holds when the subgraph induced by $\mathcal{S}$ is complete, i.e. when $\mathcal{S}$ is a clique. We refer the reader to \cite{gharesifard2017distributed} for a detailed discussion of the underlying communication graph structure. 

\section{Properties of the Objective Function}\label{sec.properties_of_objective}

While we have imposed no constraints on the objective function $f$ in Section \ref{sec.preliminaries}, many useful results regarding the greedy algorithm rely on $f$ possessing several properties most common of which are the following:
\begin{enumerate}[i )]
    \item \textit{Normalized: } $f(\emptyset) = 0$ \label{property.normalized}

    \item \textit{Monotone: } For $\mathcal{S} \subseteq \mathcal{Q} \subseteq \mathcal{V}$, $f(\mathcal{S}) \leq f(\mathcal{Q})$. \label{property.monotone}

    \item \textit{Submodular: } For $\mathcal{S} \subseteq \mathcal{Q} \subset \mathcal{V}$ and $v \in \mathcal{V} \setminus \mathcal{Q}$, the following holds: \label{property.submodular}
    \begin{equation}
        f(\mathcal{S} \cup \{v\}) - f(\mathcal{S}) \geq f(\mathcal{Q} \cup \{v\}) - f(\mathcal{Q}).
    \end{equation}
\end{enumerate}
Note that property \ref{property.submodular} (submodularity) provides a relationship between marginal rewards and is equivalently written as $\Delta(v|\mathcal{S}) \geq \Delta(v|\mathcal{Q})$.

The authors of \cite{conforti1984submodular} introduce a notion of curvature of submodular set functions that they use to unify distinct suboptimality guarantees for the greedy algorithm. Their definition of curvature, denoted by $\alpha_c$, is defined as follows.
\begin{definition}(Total Curvature \cite{conforti1984submodular})\label{def.conforti_curvature}
    Let $f$ be normalized, monotone, and submodular. Then the curvature of $f$ denoted by $\alpha_c$ is defined as
\begin{equation}\label{eq.conforti_curvature}
    \alpha_c \triangleq \underset{v \in \mathcal{V}^*}{\max} \frac{\Delta(v| \emptyset) - \Delta(v | \mathcal{V} \setminus {v})}{\Delta(v | \emptyset)}
    \end{equation}
    where $\mathcal{V}^* \triangleq \{v \in \mathcal{V}: f(v) \geq 0 \}$. 
\end{definition}

In our analysis, we utilize a more general notion of curvature given in \cite{bogunovic2018robust}.
\begin{definition}[Generalized Curvature {\cite[definition 2]{bogunovic2018robust}}]\label{def.bogunovic_curvature}
    Consider a normalized, monotone function $f: 2^{\mathcal{V}} \mapsto \R_{\geq 0}$. The generalized curvature is the smallest scalar $\alpha$ s.t.
    \begin{equation}
        \Delta(v | (\mathcal{S} \cup \mathcal{Q}) \setminus v) \geq (1 - \alpha)\Delta(v | \mathcal{S} \setminus v)
    \end{equation}
    for all $\mathcal{S}, \mathcal{Q} \subseteq \mathcal{V}$ and $v \in \mathcal{S} \setminus \mathcal{Q}$.
\end{definition}
Proposition 2 in \cite{friedrich2019greedy} gives that, for submodular, monotone functions, $\alpha$ given in Definition \ref{def.bogunovic_curvature} satisfies $\alpha \leq \alpha_c$ where $\alpha_c$ is given by Definition \ref{def.conforti_curvature}. We note that we do not expect the relationship $\alpha \leq \alpha_c$ to provide meaningful improvements over bounds utilizing $\alpha_c$. Instead, the relationship $\alpha \leq \alpha_c$ suggests that our use of Definition \ref{def.bogunovic_curvature} does not negatively influence our results with respect to bounds produced using Definition \ref{def.conforti_curvature}.

To analyze the suboptimality of the greedy algorithm given a non-submodular objective function, we utilize the definition of inverse generalized curvature presented in \cite{bogunovic2018robust}.
\begin{definition}[Inverse Generalized Curvature {\cite[definition 2]{bogunovic2018robust}}]\label{def.bogunovic_inverse_curvature}
    Consider a normalized, monotone function $f: 2^{\mathcal{V}} \mapsto \R_{\geq 0}$. The inverse generalized curvature is the smallest scalar $\beta$ s.t.
    \begin{equation}
        \Delta(v | \mathcal{S} \setminus v) \geq (1 - \beta) \Delta(v | (\mathcal{S} \setminus v) \cup \mathcal{Q})
    \end{equation}
    for all $\mathcal{S}, \mathcal{Q} \subseteq \mathcal{V}$ and $v \in \mathcal{S} \setminus \mathcal{Q}$. The function is submodular \textit{iff} $\beta = 0$ and modular \textit{iff} $\beta = \alpha = 0$. 
\end{definition}
In general $\alpha$ can be different from $\beta$.

\section{Bounds for the Greedy Algorithm Given a Non-Submodular Objective Function}\label{sec.guarantees}

The main contributions of this work are given in Theorems \ref{thm.greedy_bound} and \ref{thm.lim_inf_greedy_bound}. Throughout, we assume that the element of the solution contributed by agent $i$ or at the $i^{\text{th}}$ planning step of the greedy algorithm satisfies 
\begin{equation}\label{eq.eta_optimal}
    x_i \in \{\bar{x}_i \in \mathcal{X}_i : \eta \Delta(\bar{x}_i | x_{\mathcal{N}_i}) \geq \underset{\hat{x}_i \in \mathcal{X}_i}{\max} \Delta(\hat{x}_i | x_{\mathcal{N}_i}) \}.
\end{equation}
An element $x_i$ satisfying \eqref{eq.eta_optimal} is called $\eta$-optimal.

We note that $\eta$-optimality at each planning step is generally not guaranteed in practice. However, the consideration of $\eta$-optimality here provides insight into the effects of using planning techniques, such as Monte Carlo Tree Search, that do not necessarily provide optimal solutions, but are likely nearly optimal. For example, each agent running Monte Carlo Tree search using the d-UCT proposed in \cite{best2019dec} would allow agents to begin planning simultaneously and adapt plans as new information is gained until the planning window for each agent closes. 

\subsection{Suboptimality of the Greedy Algorithm given a Non-submodular Objective Function Under a Simple Partition Matroid Constraint}
\begin{theorem}\label{thm.greedy_bound}
Given a normalized, monotone objective function with $\alpha$ and $\beta$ given in Definitions \ref{def.bogunovic_curvature} and \ref{def.bogunovic_inverse_curvature}, and assuming $\eta$-optimal planning, the greedy algorithm guarantees a solution $x$ satisfying
    \begin{equation}\label{eq.non_sub_good_comms}
        \frac{f(x)}{f(x^*)} \geq \frac{1 - \beta}{\eta + (1 - \beta)\alpha }
    \end{equation}
\end{theorem}
The proof of Theorem \ref{thm.greedy_bound} is provided in Appendix \ref{ap.greedy}.

When $f$ is modular we have $\alpha = \beta = 0$. Thus, when $\eta = 1$ our bound reflects the optimality of the greedy algorithm proven in \cite{edmonds1971matroids}. When $\beta = 0$ and $\eta = 1$, our bound improves the $1/(1 + \alpha_c)$ bound of \cite{conforti1984submodular} by Proposition 2 of \cite{friedrich2019greedy} which gives that $\alpha \leq \alpha_c$ where $\alpha$ is given by Definition \ref{def.bogunovic_curvature}. When $\beta = 0$ and $\alpha = 1$, our bound is equivalent to that of \cite{singh2009efficient}. When $\alpha = 1$ and $\eta = 1$, our bound is equivalent to \cite{gatmiry2021network} and to the single matroid constrained case of \cite{nong2019maximize}.

\subsection{Suboptimality of the Greedy Algorithm with Limited Information given a Non-submodular Objective Function}

Theorem \ref{thm.lim_inf_greedy_bound} is inspired by distributed path planning in environments where communication is potentially unreliable. Generally, distributed execution of the greedy algorithm relies on each agent having access to the decisions of all preceding agents. Here we consider that unreliable communications have limited the information of each agent such that it must make a decision while having access to the decisions of only a subset of the preceding agents.

\begin{theorem}\label{thm.lim_inf_greedy_bound}
Given a normalized, monotone objective function with generalized curvature $\alpha$ and inverse generalized curvature $\beta$ given in Definitions \ref{def.bogunovic_curvature} and \ref{def.bogunovic_inverse_curvature}, respectively, the greedy algorithm guarantees a solution $x$ satisfying
    \begin{equation}\label{eq.general_non_sub_bound}
        \frac{f(x)}{f(x^*)} \geq \frac{(1 - \beta)^2}{(1 - \beta)^2 + (\alpha + \eta - 1 + \beta - \alpha \beta) k^*(G) }
    \end{equation}
    where $k^*(G)$ represents the fractional clique cover number of the underlying communication graph $G$.
\end{theorem}
The proof of Theorem \ref{thm.lim_inf_greedy_bound} is given in Appendix \ref{ap.lim_inf_greedy}.
We note that, due to the step in equation \eqref{eq.thm_2_intermediate_step_2} in the proof of Theorem \ref{thm.lim_inf_greedy_bound}, when $k^* = 1$ the bound in \eqref{eq.general_non_sub_bound} does not reduce to \eqref{eq.non_sub_good_comms}. For this reason, we believe a tighter bound is achievable in future work. Regardless, this result provides, to the best of the authors knowledge, the first suboptimality guarantees for the greedy algorithm with limited information given a non-submodular objective function. Notably, when $\beta = 0$ and $\alpha = \eta = 1$, we recover the approximation bound of \cite{grimsman2018impact}. Furthermore, when $\beta = 0$, the bounds of Theorems \ref{thm.lim_inf_greedy_bound} and \eqref{eq.non_sub_good_comms} each reduce to $1/(\alpha + \eta)$.

\section{The Benefit of Search Objective Function}\label{sec.benefit_of_search}

We consider a robotic search application where a team of robots seeks to cooperatively locate an unknown number of objects. We refer to the corresponding objective function as the \textit{benefit of search}. The benefit of search gives the expected reduction in risk at a location as a result of obtaining $k$ measurements of the number of objects and of the environment type in that location. 

We specifically consider the benefit of search as it addresses practical issues such as environmental influence on sensor performance, false alarms, and multiple visits to a single location where new measurements are not conditionally independent of previous measurements. These practical considerations come at the cost of the benefit of search being non-submodular such that classic approximation guarantees for submodular objective functions are not applicable. Thus, the benefit of search provides a meaningful example objective function to illustrate the value of Theorems \ref{thm.greedy_bound} and \ref{thm.lim_inf_greedy_bound}.

Note that we provide a slightly different formulation of the benefit of search from that derived in \cite{yetkin2016acquiring,mcmahon2017towards,biggs2019performance}. Specifically, our derivation does not rely on a Bayesian optimal estimate of the environment type at a location facilitating the proof of Theorem \ref{thm.bos_monotone} which gives that the benefit of search is monotone. Additionally, we show, in Theorem \ref{thm.bos_normalized}, that the benefit of search is normalized. Theorems \ref{thm.bos_normalized} and \ref{thm.bos_monotone} together ensure that the bounds of Theorems \ref{thm.greedy_bound} and \ref{thm.lim_inf_greedy_bound} apply to the benefit of search objective function. 

We assume that a search area is composed of disjoint cells. Each cell has a distinct environment type and a number of targets of interest to a team of search agents. We consider, for now, a single cell of interest that we label $h_i$. Let $\mathcal{Z}_i$ be the set of target observations and $\mathcal{Y}_i$ be the set of observations of the environment type within $h_i$. Our sensor model gives the probability of obtaining a measurement $z \in \mathcal{Z}_i$ given the true number of targets $t \in \mathcal{T}_i$ and environmental conditions $e \in \mathcal{E}_i$ and is given as 
\begin{equation}\label{eq.target_sensor_characterization}
    P(z|t,e) = \sum_{k = 0}^{\min(t,z)} {t \choose k} D_e^k (1-D_e)^{t-k} (1-A_e)A_e^{z - k}
\end{equation}
where $D_e$ gives the probability of detection and $A_e$ gives the probability of one or more false alarms. Note that both $D_e$ and $A_e$ depend on the true environment type $e$. We now drop the $i$ subscript denoting the cell $h_i$ as we will consider no other cells in the following derivations and definitions. We then have the Bayesian updates for the beliefs on the number of targets and the environmental conditions given by
\begin{align}
    P(t|z,e) &= \frac{P(z|t,e)P(t|e)}{P(z|e)}\\
    P(e|y) &= \frac{P(y|e)P(e)}{P(y)}
\end{align}
We assume that the number of targets at a location is independent of the environment type. That is, $P(t|e) = P(t)$. The posterior belief on the number of targets conditioned on environmental measurements is thus given by
\begin{equation}
    P(t|z,y) = \sum_{e \in \mathcal{E}} P(t|z,e)P(e|y).
\end{equation}

Given a measurement $z$, the team of agents must estimate the true number of targets $t$ within the cell. We consider the case that overestimating the number of targets in the cell may be preferable to the alternative. As such, we impose a cost on the estimate $\delta(z) \in \mathcal{T}$ that handles this consideration.
\begin{equation}
    L(t, \delta(z)) = c_i|t - \delta(z)|\quad \text{for }i \in \{1,2 \}
\end{equation}
The posterior expected loss (risk) of computing the estimate $\delta(z)$ is then given by
\begin{equation}\label{eq.expected_loss}
    \mathbb{E}[L(t,\delta(z))|z,y] = \sum_{t \in \mathcal{T}} P(t|z,y)L(t, \delta(z)).
\end{equation}
and, given no measurements, the posterior expected loss given an estimate $\delta \in \mathcal{T}$ is given as
\begin{equation}
    \mathbb{E}[L(t,\delta)] = \sum_{t \in \mathcal{T}} P(t)L(t, \delta)
\end{equation}
with the Bayes estimate found as
\begin{equation}
    \delta^* = \underset{\delta \in \mathcal{T}}{\arg \min} \mathbb{E}[L(t,\delta)].
\end{equation}

Note that \eqref{eq.expected_loss} represents the primary deviation of our work from the benefit of search as described in \cite{mcmahon2017towards}. Specifically, we compute risk using $P(t|z,y)$ instead of $P(t|z,e)$. As a result of this deviation, we are not required to find an optimal estimate of $e$ in order to compute anticipated risk.

The current risk is defined using the Bayes estimate and is given as
\begin{equation}\label{eq.current_risk}
    r(0) = \mathbb{E}[L(t,\delta^*)].
\end{equation}
The Bayes estimate of the number of targets given measurements $z$ and $y$ is given as
\begin{equation}
    \delta^*(z) = \underset{\delta(z) \in \mathcal{T}}{\arg \min} \mathbb{E}[L(t,\delta(z))|z,y]
\end{equation}
and the anticipated risk conditioned on measurements $z$ and $y$ is
\begin{equation}\label{eq.risk_cond_z_y}
    r(z,y) = \mathbb{E}[L(t, \delta^*(z))|z,y].
\end{equation}

Naturally, we do not have measurements $z$ and $y$ when planning. Therefore, we find the expectation of \eqref{eq.risk_cond_z_y} over the space of possible measurements to get the anticipated risk
\begin{equation}
    \mathbb{E}[r(z, y)] = \sum_{z \in \mathcal{Z}} \sum_{y \in \mathcal{Y}} P(z,y) \mathbb{E}[L(t,\delta^*(z))|z,y].
\end{equation}
The joint probability $P(z,y)$ of obtaining measurements $z$ and $y$ is
\begin{equation}
    P(z,y) = \sum_{t \in \mathcal{T}} \sum_{e \in \mathcal{E}} P(z|t,e) P(y | e) P(t) P(e).
\end{equation}
Because we use new measurements to update our prior distributions $P(t)$ and $P(e)$, the conditional probability of a second set of measurements is 
\begin{align}
    P(z_2, y_2| z_1, y_1) &= \sum_{t \in \mathcal{T}} \sum_{e \in \mathcal{E}} P(z_2|t,e) P(y_2 | e) \nonumber \\ \quad & \times P(t|z_1, y_1) P(e|y_1).
\end{align}
Note that $P(z_1, z_2, y_1, y_2) = P(z_2, y_2 | z_1, y_1)P(z_1, y_1)$. Expanding upon this, we see that for $k$ measurements we have
\begin{align}
    P(z, y) &= \sum_{t \in \mathcal{T}} \sum_{e \in \mathcal{E}} P(z_k|t,e) P(y_k | e)  \nonumber \\ \quad & \times P(t|z_1 , \ldots, z_{k-1}, y_1, \ldots, y_{k-1}) \nonumber \\ \quad & \times P(e|y_1, \ldots, y_{k-1}) \nonumber \\ \quad & \times  P(z_1, \ldots, z_{k-1}, y_1, \ldots, y_{k-1})
\end{align}
where $z = \{z_1, \ldots, z_k \}$ and $y = \{ y_1, \ldots, y_k \}$ and $P(z,y) = P(z_1, \ldots,  z_k, y_1, \ldots, y_k)$. The anticipated risk given $k$ measurements is then found as
\begin{equation}\label{eq.anticipated_risk}
    r(k) = \sum_{z \in \mathcal{Z}^k} \sum_{y \in \mathcal{Y}^k} P(z, y)  \mathbb{E}[L(t,\delta^*(z))|z,y].
\end{equation}

We use equations \eqref{eq.current_risk} and \eqref{eq.anticipated_risk} to define the benefit of searching cell $h_i$ $k$ times as
\begin{equation}\label{eq.benefit_of_search}
    f_i(k) \triangleq r(0) - r(k).
\end{equation}

\begin{figure}
\centering
\begin{subfigure}{.25\columnwidth}
    \centering
    \includegraphics[width=0.98\textwidth]{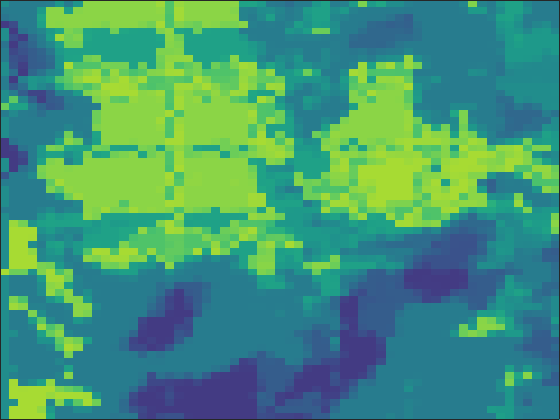}
\end{subfigure}%
\begin{subfigure}{.25\columnwidth}
    \centering
    \includegraphics[width=0.98\textwidth]{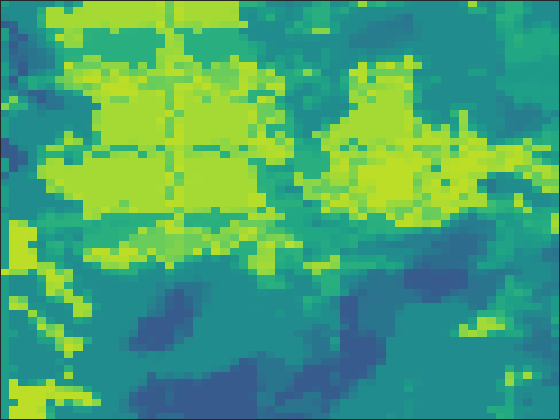}
\end{subfigure}%
\begin{subfigure}{.25\columnwidth}
    \centering
    \includegraphics[width=0.98\textwidth]{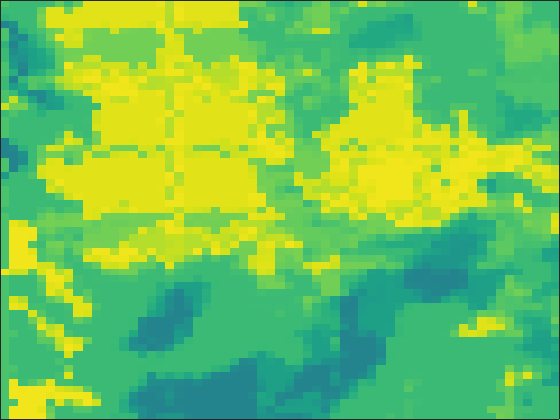}
\end{subfigure}%
\begin{subfigure}{.25\columnwidth}
    \centering
    \includegraphics[width=0.98\textwidth]{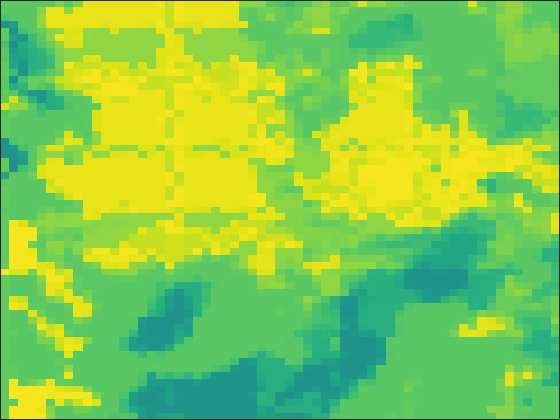}
\end{subfigure}%

\vspace{0.08em}

\begin{subfigure}{.25\columnwidth}
    \centering
    \includegraphics[width=0.98\textwidth]{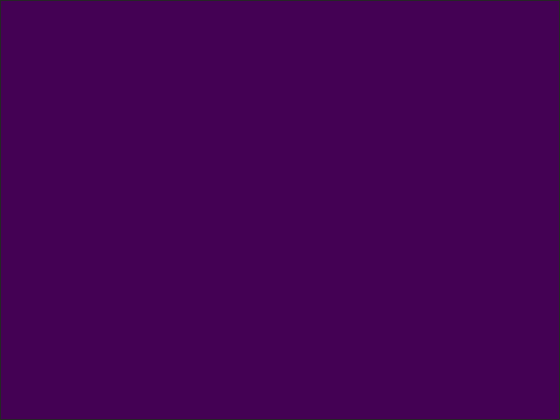}
\end{subfigure}%
\begin{subfigure}{.25\columnwidth}
    \centering
    \includegraphics[width=0.98\textwidth]{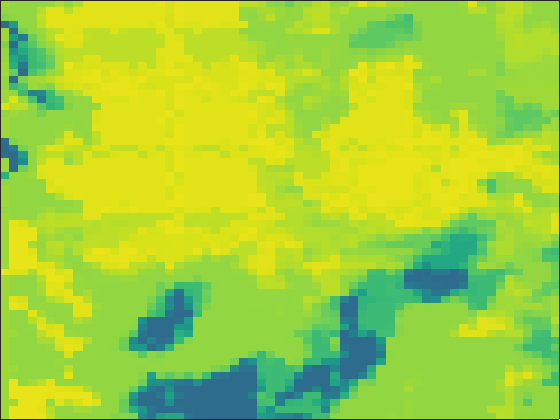}
\end{subfigure}%
\begin{subfigure}{.25\columnwidth}
    \centering
    \includegraphics[width=0.98\textwidth]{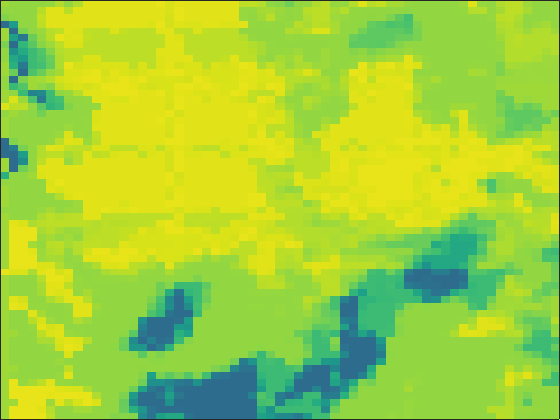}
\end{subfigure}%
\begin{subfigure}{.25\columnwidth}
    \centering
    \includegraphics[width=0.98\textwidth]{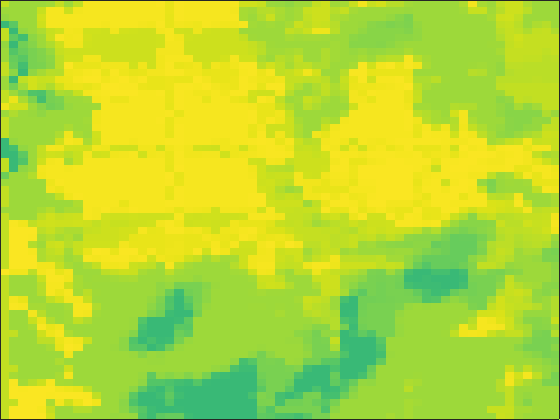}
\end{subfigure}%

\vspace{0.1em}

\begin{subfigure}{.25\columnwidth}
    \centering
    \includegraphics[width=0.98\textwidth]{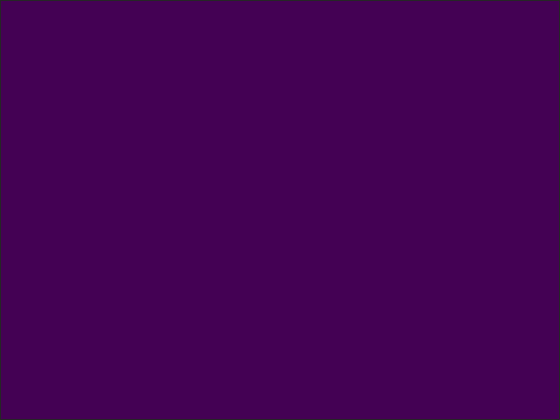}
    \caption*{$k = 1$}
\end{subfigure}%
\begin{subfigure}{.25\columnwidth}
    \centering
    \includegraphics[width=0.98\textwidth]{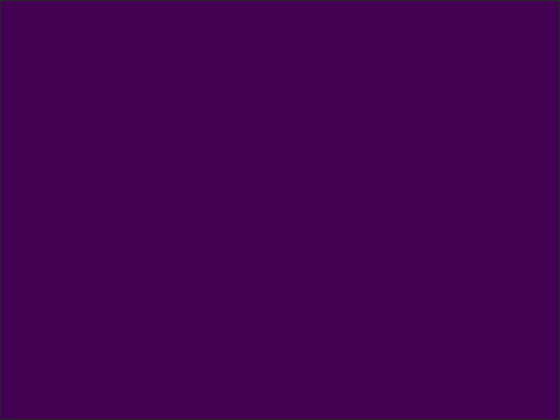}
    \caption*{$k = 2$}
\end{subfigure}%
\begin{subfigure}{.25\columnwidth}
    \centering
    \includegraphics[width=0.98\textwidth]{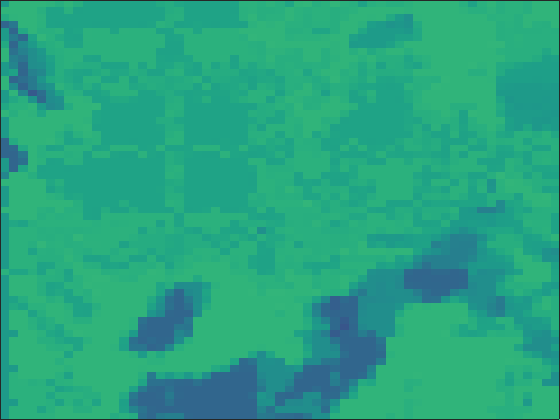}
    \caption*{$k = 3$}
\end{subfigure}%
\begin{subfigure}{.25\columnwidth}
    \centering
    \includegraphics[width=0.98\textwidth]{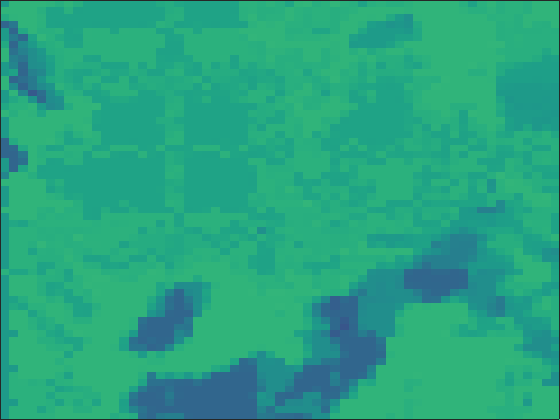}
    \caption*{$k = 4$}
\end{subfigure}%

\caption{The first row gives the normalized anticipated reward for $k$ visits to each cell. The second row gives the generalized curvature from Definition \ref{def.bogunovic_curvature} for each cell assuming at most $k$ visits to each cell. The third row gives the inverse generalized curvature from Definition \ref{def.bogunovic_inverse_curvature} for each cell assuming at most $k$ visits to each cell. All values are between 0 and 1 with darker regions indicating small values and lighter regions indicating large values.}
\label{fig.boston_figures}
\end{figure}

\section{Properties of the Benefit of Search}\label{sec.benefit_of_search_properties}
To show that the benefit of search is an appropriate objective function given the conditions of Theorems \ref{thm.greedy_bound} and \ref{thm.lim_inf_greedy_bound}, we first show that the benefit of search is normalized and monotone. Furthermore, we characterize the parameters $\alpha$ and $\beta$ for each cell in a discrete map where the environment type prior probability distribution $P(e)$ comes from real-world data that was acquired by an autonomous underwater vehicle during a subsea survey of Boston Harbor.

\begin{theorem}[Normalized]\label{thm.bos_normalized}
    The benefit of search as described in Section \ref{sec.benefit_of_search} is normalized.
\end{theorem}
\begin{IEEEproof}
    Clearly, if no new measurements are taken then the anticipated risk given no measurements is simply the current risk once again. Therefore,
    \begin{equation}
        f(0) = r(0) - r(0) = 0.
    \end{equation}
\end{IEEEproof}

\begin{theorem}[Monotone]\label{thm.bos_monotone}
    The benefit of search as described in Section \ref{sec.benefit_of_search} is monotone.
\end{theorem}
The proof of Theorem \ref{thm.bos_monotone} is given in Appendix \ref{ap.monotone}.

In order to characterize the parameters $\alpha$ and $\beta$, we consider that there are 3 possible environment types. For simplicity, we assume that all search agents have the same sensors and therefore $P(z|t,e)$ and $P(y|e)$ is the same across all agents. Specifically, we use 
\begin{equation}
    P(y|e) = \begin{pmatrix} 0.82 & 0.09 & 0.09 \\ 0.08 & 0.84 & 0.08 \\ 0.06 & 0.06 & 0.88 \end{pmatrix}
\end{equation}
where the $ij^{\text{th}}$ element of $P(y|e)$ gives the probability that the environmental sensor will measure the environment type to be $y = j$ given the true environment type is $e = i$. In constructing $P(z|t,e)$, we use $D_e = ( 0.65, 0.8, 0.95 )$ and $A_e = (0.4, 0.3, 0.05)$.

We assume $t \in \{0, 1, 2\}$ with prior probability distribution $P(t)$ chosen to be a truncated Poisson distribution meaning that most locations are considered unlikely to contain an object of interest. The cost of underestimating the number of targets is $c_1 = 3$ and the cost of overestimating the number of targets is $c_2 = 1$.  The normalized anticipated reward as well as the generalized curvature $\alpha$ and inverse generalized curvature $\beta$ for each cell in the search area  corresponding to a fixed maximum number of search passes $k$ are shown in Figure \ref{fig.boston_figures}. 

In the specific example shown in Figure \ref{fig.boston_figures}, $\alpha$ increases significantly in the second column reflecting a relatively small increase in reward for visiting each cell a second time as compared to the reward gained upon a single search pass. The increased value of $\beta$ in the third column corresponds to an increased marginal reward upon a third visit to each cell as compared to the marginal reward attained upon a second visit to each cell. The maximum values of $\alpha$ and $\beta$ corresponding to each value of $k$ are given in Table \ref{table.curvature_values}.

Note that in Definitions \ref{def.bogunovic_curvature} and \ref{def.bogunovic_inverse_curvature}, $\alpha$ and $\beta$ are characterized based on the ground set of a matroid. Because there is no limit on how many times a single cell may be visited, we suppose that the marginal gain for visiting a cell more than $k$ times is zero. This results in $\alpha = 1$ for each cell and $\beta$ is the maximum value over all cells for the maximum value of $k$ considered before fixing additional marginal gains to be zero. We can, therefore bound the value of a joint path planned using the greedy algorithm using $\alpha = 1$ and $\beta$ is the value given in Table \ref{table.curvature_values} for the maximum value of $k$ considered. Thus, for $k = 4$ and $\eta = 1.25$, Theorem \ref{thm.greedy_bound} guarantees an approximation ratio of $(1 - 0.6564)/(1.25 + (1 - 0.6564)) = 0.2156$. The approximation guarantee of Theorem \ref{thm.lim_inf_greedy_bound} may be calculated likewise.

\begin{table}
\normalsize
\centering
\begin{tabular}{|c|c|c|c|c|}
    \hline
     & k = 1 & k = 2 & k = 3 & k = 4 \\ 
     \hline
     $\underset{h_i}{\max} \ \alpha$ & 0 & 0.9688 & 0.9688 & 0.9923 \\
     \hline
     $\underset{h_i}{\max} \ \beta$ & 0 & 0 & 0.6564 & 0.6564 \\
     \hline
\end{tabular}
\caption{Maximum values of $\alpha$ and $\beta$ with respect to the maximum number of allowed search passes $k$.}
\label{table.curvature_values}
\end{table}

\section{CONCLUSIONS}\label{sec.conclusions}
We present worst case performance bounds for the greedy algorithm in seeking to maximize a normalized, monotone, but not necessarily submodular objective function under a simple partition matroid constraint. We further provide bounds on the performance of the greedy algorithm assuming limited information at each planning step. We demonstrate that a variant of the benefit of search objective function is normalized and monotone, but not submodular. We characterize the curvature of the benefit of search using real-world data collected by an autonomous underwater vehicle. We show that theoretical approximation guarantees are achievable despite non-submodularity of the objective function.

\section*{APPENDICES}
\appendices

\section{Proof of Theorem \ref{thm.greedy_bound}} \label{ap.greedy}
\begin{IEEEproof}
    \begin{align}
        f(x^*) &= f(x) \nonumber \\ & \quad + \sum_{i \in \mathcal{A}} \Delta(x^*_i | x^*_{1:i-1}, x) - \sum_{i \in \mathcal{A}}\Delta(x_i | x_{1:i-1}, x^*) \\
               &= \sum_{i \in \mathcal{A}} \Delta(x_i | x_{1:i-1}) \nonumber \\ & \quad + \sum_{i \in \mathcal{A}} \Delta(x^*_i | x^*_{1:i-1}, x) - \sum_{i \in \mathcal{A}} \Delta(x_i | x_{1:i-1}, x^*) \\
               &= \sum_{i \in \mathcal{A}} \Delta(x^*_i | x^*_{1:i-1}, x) \nonumber \\ & \quad + \sum_{i \in \mathcal{A}} \left[ \Delta(x_i | x_{1:i-1}) - \Delta(x_i | x_{1:i-1}, x^*) \right] \\ \label{eq.non_sub_good_comms_1}
               &\leq \sum_{i \in \mathcal{A}} \Delta(x^*_i | x^*_{1:i-1}, x) + \alpha \sum_{i \in \mathcal{A}} \Delta(x_i | x_{1:i-1}) 
    \end{align}
    where the inequality of \eqref{eq.non_sub_good_comms_1} is by Definition \ref{def.bogunovic_curvature}.
    
    Multiplying both sides of \eqref{eq.non_sub_good_comms_1} by $1 - \beta$ gives
    \begin{align}
        (1 - \beta) f(x^*) &\leq (1 - \beta) \sum_{i \in \mathcal{A}} \Delta(x^*_i | x^*_{1:i-1}, x) \nonumber \\ & \quad + (1 - \beta) \alpha \sum_{i \in \mathcal{A}} \Delta(x_i | x_{1:i-1}) \\
                           &\leq \sum_{i \in \mathcal{A}} \Delta(x^*_i | x_{1:i-1}) \nonumber \\ & \quad + (1 - \beta) \alpha \sum_{i \in \mathcal{A}} \Delta(x_i | x_{1:i-1}) \\
                           &\leq \eta \sum_{i \in \mathcal{A}} \Delta(x_i | x_{1:i-1}) \nonumber \\ & \quad + (1 - \beta) \alpha \sum_{i \in \mathcal{A}} \Delta(x_i | x_{1:i-1}) \\
                           &= \eta f(x) + (1 - \beta) \alpha f(x)
    \end{align}
    where the second inequality is by Definition \ref{def.bogunovic_inverse_curvature} and the third inequality is by $\eta$-optimality of the greedy selection.
\end{IEEEproof}

\section{Proof of Theorem \ref{thm.lim_inf_greedy_bound}} \label{ap.lim_inf_greedy}
\begin{IEEEproof}
    \begin{equation}
        f(x^*) = f(x) + \sum_{i \in \mathcal{A}} \Delta(x^*_i | x^*_{1:i-1}, x) - \sum_{i \in \mathcal{A}}\Delta(x_i | x_{1:i-1}, x^*).
    \end{equation}
    Multiplying both sides by $1 - \beta$ gives
    \begin{align}
        (1 - \beta) f(x^*) &= (1 - \beta) f(x) \nonumber \\ & \quad + (1 - \beta) \sum_{i \in \mathcal{A}} \Delta(x^*_i | x^*_{1:i-1}, x) \nonumber \\ & \quad - (1 - \beta) \sum_{i \in \mathcal{A}} \Delta(x_i | x_{1:i-1}, x^*).
    \end{align}
    By Definition \ref{def.bogunovic_inverse_curvature}
    \begin{equation}
        (1 - \beta) \sum_{i \in \mathcal{A}} \Delta(x^*_i | x^*_{1:i-1}, x) \leq \sum_{i \in \mathcal{A}} \Delta(x^*_i | x_{\mathcal{N}_i}).
    \end{equation}
    and by $\eta$ optimality $\Delta(x^*_i | x_{\mathcal{N}_i}) \leq \eta \Delta(x_i | x_{\mathcal{N}_i})$ such that
    \begin{align}
        (1 - \beta) f(x^*) &\leq (1 - \beta) f(x) \nonumber \\ & \quad + \eta \sum_{i \in \mathcal{A}} \Delta(x_i | x_{\mathcal{N}_i}) \nonumber \\ & \quad - (1 - \beta) \sum_{i \in \mathcal{A}} \Delta(x_i | x_{1:i-1}, x^*).
    \end{align}
    Adding and subtracting $(1 - \beta) \sum_{i \in \mathcal{A}} \Delta(x_i | x_{\mathcal{N}_i})$ to the right-hand side and combining summations and like terms gives
    \begin{align}
        (1 - \beta) f(x^*) &\leq (1 - \beta) f(x) \nonumber \\ & + (\eta - (1 - \beta))\sum_{i \in \mathcal{A}} \Delta(x_i | x_{\mathcal{N}_i}) \nonumber \\ & + (1 - \beta) \sum_{i \in \mathcal{A}} \left[ \Delta(x_i | x_{\mathcal{N}_i}) - \Delta(x_i | x_{1:i-1}, x^*) \right].
    \end{align}
    By Definition \ref{def.bogunovic_curvature}
    \begin{align}
        (1 - \beta) f(x^*) &\leq (1 - \beta) f(x) \nonumber \\ & \quad + (\eta - (1 - \beta))\sum_{i \in \mathcal{A}} \Delta(x_i | x_{\mathcal{N}_i}) \nonumber \\ & \quad + (1 - \beta) \alpha \sum_{i \in \mathcal{A}} \Delta(x_i | x_{\mathcal{N}_i})
    \end{align}
    which simplifies to
    \begin{align} \label{eq.thm_2_intermediate_step_1}
        (1 - \beta) f(x^*) &\leq (1 - \beta) f(x) \nonumber \\ & \quad + (\alpha + \eta - 1 + \beta - \alpha \beta)\sum_{i \in \mathcal{A}} \Delta(x_i | x_{\mathcal{N}_i}).
    \end{align}
    
    The remainder of the proof is inspired by the proof of Theorem 1 in \cite{grimsman2018impact} to which we refer the reader for greater intuition regarding set of scalars $\{ y_c\}_{c \in \mathcal{K}(G)}$. 
    
    Suppose that we have a set of scalars $\{ y_c\}_{c \in \mathcal{K}(G)}$ where $\mathcal{K}(G)$ is the set of all cliques (fully connected subsets of $\mathcal{A}$) such that $y_c \geq 0$ for all $c$ and $\sum_{c \in \mathcal{K}(G):i \in c}y_c \geq 1$ for all $i$. Then
    \begin{align}
        \label{eq.multiply_sum_of_scalars}
        \sum_{i \in \mathcal{A}} \Delta(x_i | x_{\mathcal{N}_i}) &\leq \sum_{i \in \mathcal{A}} \Delta(x_i | x_{\mathcal{N}_i}) \left[ \sum_{c:i \in c}y_c \right]    \\
               &= \sum_{i \in \mathcal{A}} \sum_{c:i \in c} y_c \Delta(x_i | x_{\mathcal{N}_i})                     \\ \label{eq.sum_over_cliques}
               &= \sum_{c \in \mathcal{K}(G)} y_c \sum_{i \in c}  \Delta(x_i | x_{\mathcal{N}_i}).
    \end{align}
    Therefore \eqref{eq.thm_2_intermediate_step_1} and \eqref{eq.sum_over_cliques} give
    \begin{align}
        (1 - \beta) f(x^*) &\leq (1 - \beta) f(x) \nonumber \\ & + (\alpha + \eta - 1 + \beta - \alpha \beta) \nonumber \\ & \quad \times \sum_{c \in \mathcal{K}(G)} y_c \sum_{i \in c}  \Delta(x_i | x_{\mathcal{N}_i}).
    \end{align}
    Multiplying both sides by $1 - \beta$ once more, we have
    \begin{align}\label{eq.thm_2_intermediate_step_2}
        (1 - \beta)^2 f(x^*) &\leq (1 - \beta)^2 f(x) \nonumber \\ & \quad + (\alpha + \eta - 1 + \beta - \alpha \beta) \nonumber \\ & \quad \times \sum_{c \in \mathcal{K}(G)} y_c (1 - \beta) \sum_{i \in c}  \Delta(x_i | x_{\mathcal{N}_i})         \\
    \end{align}
    where 
    \begin{align}
        \label{eq.submodular_ratio}
              \sum_{c \in \mathcal{K}(G)} y_c (1 - \beta) \sum_{i \in c}  \Delta(x_i | x_{\mathcal{N}_i}) &\leq  \sum_{c \in \mathcal{K}(G)} y_c \sum_{i \in c} \Delta(x_i | x_{\mathcal{N}_i \cap c})         \\
        \label{eq.clique_sum}
               &= \sum_{c \in \mathcal{K}(G)} y_c f(x_c)                                                              \\
        \label{eq.monotone}
               & \leq \sum_{c \in \mathcal{K}(G)} y_c f(x).
    \end{align}
    where \eqref{eq.submodular_ratio} is by Definition \ref{def.bogunovic_inverse_curvature}, \eqref{eq.clique_sum} is by Definition \ref{def.marginal_reward}, and \eqref{eq.monotone} is by the monotone property of $f$. From \eqref{eq.thm_2_intermediate_step_2} and \eqref{eq.monotone} we have
    \begin{align}
        (1 - \beta)^2 f(x^*) &\leq (1 - \beta)^2 f(x) \nonumber \\ & \quad + (\alpha + \eta - 1 + \beta - \alpha \beta) \sum_{c \in \mathcal{K}(G)} y_c f(x)
    \end{align}

    The performance bound follows as
    \begin{equation}\label{ineq.bound.2}
        \frac{f(x)}{f(x^*)} \geq \frac{(1 - \beta)^2}{(1 - \beta)^2 + (\alpha + \eta - 1 + \beta - \alpha \beta) \sum_{c \in \mathcal{K}(G)} y_c }
    \end{equation}
    To make the bound in \eqref{ineq.bound.2} as tight as possible, one can solve the following optimization:
    \begin{equation}
        \underset{y}{\min} \sum_{c \in K(G)}y_c
    \end{equation}
    subject to $\sum_{c \in K(G): i \in c}y_c \geq 1, \ \forall i$ and $y_c \geq 0, \ \forall c$.
    The solution to this optimization problem is the fractional clique cover number of the communication graph $G$ and is denoted by $k^*(G)$. Thus our final result is given as
    \begin{equation}
        \frac{f(x)}{f(x^*)} \geq \frac{(1 - \beta)^2}{(1 - \beta)^2 + (\alpha + \eta - 1 + \beta - \alpha \beta) k^*(G) }
    \end{equation}

\end{IEEEproof}

\section{Proof of Theorem \ref{thm.bos_monotone}} \label{ap.monotone}
\begin{IEEEproof}
    To be monotone, it must be true that increasing the number of measurements increases the benefit of search. That is, we wish to show that $f(k) \geq f(k-1)$ for all $k \geq 1$. Naturally, because risk is a positive number, and the benefit of search is the difference in risk, we simply need to show that the risk is monotonically decreasing in order to prove that $f$ is monotone increasing. 

    Therefore, we seek to show that $r(k) \leq r(k-1)$. Let $z = \{z_1, \ldots, z_k \}$ and $y = \{y_1, \ldots, y_k \}$. Also, let $z^- = \{z_1, \ldots, z_{k-1} \}$ and $y^- = \{y_1, \ldots, y_{k-1} \}$. As before, we have 
    \begin{equation}
        r(k) = \sum_{z \in \mathcal{Z}^k} \sum_{y \in \mathcal{Y}^k} P(z, y)  \left [ \sum_{t \in \mathcal{T}} P(t|z,y) L(t, \delta^*(z)) \right ]
    \end{equation}
    where $\delta^*(z)$ minimizes $\sum_{t \in \mathcal{T}} P(t|z,y) L(t, \delta(z))$ such that $\delta^*(z^-)$ minimizing $\sum_{t \in \mathcal{T}} P(t|z^-,y^-) L(t, \delta(z^-))$ yields an upper bound 
    \begin{align}
        r(k) &\leq \sum_{z \in \mathcal{Z}^k} \sum_{y \in \mathcal{Y}^k} P(z, y)  \left [ \sum_{t \in \mathcal{T}} P(t|z,y) L(t, \delta^*(z^-)) \right ] \\
            &= \sum_{z^- \in \mathcal{Z}^{k-1}} \sum_{y^- \in \mathcal{Y}^{k-1}} \sum_{z_k \in \mathcal{Z}} \sum_{y_k \in \mathcal{Y}} \nonumber \\ & \quad \times P(z_k, y_k | z^-, y^-) P(z^-, y^-) \nonumber \\ & \quad \times  \left [ \sum_{t \in \mathcal{T}} P(t|z,y) L(t, \delta^*(z^-)) \right ] \\
            &= \sum_{z^- \in \mathcal{Z}^{k-1}} \sum_{y^- \in \mathcal{Y}^{k-1}} P(z^-, y^-) \nonumber \times \\ &\left [ \sum_{t \in \mathcal{T}} \sum_{z_k \in \mathcal{Z}} \sum_{y_k \in \mathcal{Y}}  P(z_k, y_k | z^-, y^-) P(t|z,y) L(t, \delta^*(z^-)) \right ] \\
            &= \sum_{z^- \in \mathcal{Z}^{k-1}} \sum_{y^- \in \mathcal{Y}^{k-1}} P(z^-, y^-) \nonumber \\ & \quad \times \left [ \sum_{t \in \mathcal{T}} \sum_{z_k \in \mathcal{Z}} \sum_{y_k \in \mathcal{Y}} P(t, z_k, y_k|z^-,y^-)L(t, \delta^*(z^-)) \right ] \\
            &= \sum_{z^- \in \mathcal{Z}^{k-1}} \sum_{y^- \in \mathcal{Y}^{k-1}} P(z^-, y^-) \nonumber \\ & \quad \times \left [ \sum_{t \in \mathcal{T}} P(t |z^-,y^-)L(t, \delta^*(z^-)) \right ] \\ 
            &= r(k-1) 
    \end{align}
\end{IEEEproof}

\bibliography{sources} 

\begin{thebibliography}{10}

\bibitem{conforti1984submodular}
M.~Conforti and G.~Cornu{\'e}jols, ``Submodular set functions, matroids and the
  greedy algorithm: tight worst-case bounds and some generalizations of the
  rado-edmonds theorem,'' {\em Discrete applied mathematics}, vol.~7, no.~3,
  pp.~251--274, 1984.

\bibitem{edmonds1971matroids}
J.~Edmonds, ``Matroids and the greedy algorithm,'' {\em Mathematical
  programming}, vol.~1, no.~1, pp.~127--136, 1971.

\bibitem{fisher1978analysis}
M.~L. Fisher, G.~L. Nemhauser, and L.~A. Wolsey, ``An analysis of
  approximations for maximizing submodular set functions—ii,'' in {\em
  Polyhedral combinatorics}, pp.~73--87, Springer, 1978.

\bibitem{nemhauser1978analysis}
G.~L. Nemhauser, L.~A. Wolsey, and M.~L. Fisher, ``An analysis of
  approximations for maximizing submodular set functions—i,'' {\em
  Mathematical programming}, vol.~14, no.~1, pp.~265--294, 1978.

\bibitem{krause2012near}
A.~Krause and C.~E. Guestrin, ``Near-optimal nonmyopic value of information in
  graphical models,'' {\em arXiv preprint arXiv:1207.1394}, 2012.

\bibitem{lu2022maximizing}
C.~Lu, W.~Yang, R.~Yang, and S.~Gao, ``Maximizing a non-decreasing
  non-submodular function subject to various types of constraints,'' {\em
  Journal of Global Optimization}, pp.~1--25, 2022.

\bibitem{gatmiry2021network}
K.~Gatmiry and M.~Gomez-Rodriguez, ``The network visibility problem,'' {\em ACM
  Transactions on Information Systems (TOIS)}, vol.~40, no.~2, pp.~1--42, 2021.

\bibitem{gharesifard2017distributed}
B.~Gharesifard and S.~L. Smith, ``Distributed submodular maximization with
  limited information,'' {\em IEEE transactions on control of network systems},
  vol.~5, no.~4, pp.~1635--1645, 2017.

\bibitem{grimsman2018impact}
D.~Grimsman, M.~S. Ali, J.~P. Hespanha, and J.~R. Marden, ``The impact of
  information in distributed submodular maximization,'' {\em IEEE Transactions
  on Control of Network Systems}, vol.~6, no.~4, pp.~1334--1343, 2018.

\bibitem{singh2009efficient}
A.~Singh, A.~Krause, C.~Guestrin, and W.~J. Kaiser, ``Efficient informative
  sensing using multiple robots,'' {\em Journal of Artificial Intelligence
  Research}, vol.~34, pp.~707--755, 2009.

\bibitem{nong2019maximize}
Q.~Nong, T.~Sun, S.~Gong, Q.~Fang, D.~Du, and X.~Shao, ``Maximize a monotone
  function with a generic submodularity ratio,'' in {\em International
  Conference on Algorithmic Applications in Management}, pp.~249--260,
  Springer, 2019.

\bibitem{bai2018greed}
W.~Bai and J.~Bilmes, ``Greed is still good: maximizing monotone submodular+
  supermodular (bp) functions,'' in {\em International Conference on Machine
  Learning}, pp.~304--313, PMLR, 2018.

\bibitem{schrijver2003combinatorial}
A.~Schrijver, {\em Combinatorial optimization: polyhedra and efficiency},
  vol.~24.
\newblock Springer Science \& Business Media, 2003.

\bibitem{bian2017guarantees}
A.~A. Bian, J.~M. Buhmann, A.~Krause, and S.~Tschiatschek, ``Guarantees for
  greedy maximization of non-submodular functions with applications,'' in {\em
  International conference on machine learning}, pp.~498--507, PMLR, 2017.

\bibitem{corah2018distributed}
M.~Corah and N.~Michael, ``Distributed submodular maximization on partition
  matroids for planning on large sensor networks,'' in {\em 2018 IEEE
  Conference on Decision and Control (CDC)}, pp.~6792--6799, IEEE, 2018.

\bibitem{corah2019distributed}
M.~Corah and N.~Michael, ``Distributed matroid-constrained submodular
  maximization for multi-robot exploration: Theory and practice,'' {\em
  Autonomous Robots}, vol.~43, no.~2, pp.~485--501, 2019.

\bibitem{friedrich2019greedy}
T.~Friedrich, A.~G{\"o}bel, F.~Neumann, F.~Quinzan, and R.~Rothenberger,
  ``Greedy maximization of functions with bounded curvature under partition
  matroid constraints,'' in {\em Proceedings of the AAAI Conference on
  Artificial Intelligence}, vol.~33, pp.~2272--2279, 2019.

\bibitem{bogunovic2018robust}
I.~Bogunovic, J.~Zhao, and V.~Cevher, ``Robust maximization of non-submodular
  objectives,'' in {\em International Conference on Artificial Intelligence and
  Statistics}, pp.~890--899, PMLR, 2018.

\bibitem{best2019dec}
G.~Best, O.~M. Cliff, T.~Patten, R.~R. Mettu, and R.~Fitch, ``Dec-mcts:
  Decentralized planning for multi-robot active perception,'' {\em The
  International Journal of Robotics Research}, vol.~38, no.~2-3, pp.~316--337,
  2019.

\bibitem{yetkin2016acquiring}
H.~Yetkin, C.~Lutz, and D.~Stilwell, ``Acquiring environmental information
  yields better anticipated search performance,'' in {\em OCEANS 2016 MTS/IEEE
  Monterey}, pp.~1--6, IEEE, 2016.

\bibitem{mcmahon2017towards}
J.~McMahon, H.~Yetkin, A.~Wolek, Z.~J. Waters, and D.~J. Stilwell, ``Towards
  real-time search planning in subsea environments,'' in {\em 2017 IEEE/RSJ
  International Conference on Intelligent Robots and Systems (IROS)},
  pp.~87--94, IEEE, 2017.

\bibitem{biggs2019performance}
B.~Biggs, D.~J. Stilwell, H.~Yetkin, and J.~McMahon, ``Performance guarantees
  for receding horizon search with terminal cost,'' in {\em 2019 IEEE/RSJ
  International Conference on Intelligent Robots and Systems (iROS)},
  pp.~6362--6368, IEEE, 2019.

\end{thebibliography}
\bibliographystyle{ieeetr}


\end{document}